\documentclass[twocolumn,showpacs,preprintnumbers,amsmath,amssymb]{revtex4}

\usepackage{graphicx}   
\usepackage{epsfig}
\usepackage{psfig}
\usepackage{keyval}
\usepackage{amsmath, amsthm}
\usepackage{dcolumn}    
\usepackage{bm}         

\def\grad{\nabla}

\begin{document}



\title{Interfacial Slip in Sheared Polymer Blends}

\author{Tak Shing Lo$^1$, Maja Mihajlovic$^2$,
\footnote{Corresponding author. E-mail: shnidman@mail.csi.cuny.edu}Yitzhak
Shnidman$^{3,5}$, Wentao Li$^4$ and Dilip Gersappe$^{4,5}$}

\affiliation{$^1$The Levich Institute, City College of CUNY, New York,
NY 10031                                           \\
$^2$Department of Chemistry, City College of CUNY, New York,
NY 10031                                           \\
$^3$Department of Engineering Science and Physics, College of Staten Island 
of CUNY, Staten Island, NY 10314                   \\
$^4$Department of Materials Science and Engineering, SUNY, Stony Brook,
NY 11794                                           \\
$^5$NSF MRSEC on Polymers at Engineered Interfaces}

\date{\today} 

\begin{abstract}
We have developed a dynamic self-consistent field theory, without any
adjustable parameters, for unentangled polymer blends under shear. Our model
accounts for the interaction between polymers, and enables one to compute
the evolution of the local rheology, microstructure and the conformations of
the polymer chains under shear self-consistently. We use this model to study
the interfacial dynamics in sheared polymer blends and make a quantitative
comparison between this model and Molecular Dynamics simulations. We find
good agreement between the two methods.
\end{abstract}

\pacs{83.80.Tc, 83.50.Lh, 83.10.Rs} 
\maketitle

Self-consistent field (SCF) theory has been very useful for modeling and 
understanding the equilibrium properties of inhomogeneous polymer fluids. In 
lattice SCF theories, Markovian statistics for chains represented by lattice 
random walks are coupled with a mean field approximation for the free energy 
to determine variation of equilibrium properties across interfaces \cite{SF}. 
In contrast, interfacial dynamics and rheology of inhomogeneous polymer 
fluids driven out of equilibrium by applied stresses is still not well 
understood. Since experimental evidence indicates that the interfacial 
properties are critical to the evolution of the morphology of polymer blends, 
understanding the interplay between the interfacial dynamics and the polymer 
conformation at the interface is of great fundamental importance, and 
efficient computational models of these systems are essential for many 
industrial and biological applications.

We have recently developed a novel dynamic self-consistent field (DSCF) 
theory \cite{DSCF1} by combining the lattice random walk formalism of 
Scheutjens and Fleer's static SCF model \cite{SF} and a convective-diffusive 
lattice gas model for simple compressible fluids \cite{CDLG}. As in static 
SCF theory, the distribution of free segments interacting with a 
self-consistent field representing adjacent connected segments and walls is 
approximated by a product of free segment (one-body) probabilities. The 
effect of flow on ideal chain conformations is modeled with FENE-P dumbbells, 
and related to stepping probabilities in a random walk. Free segment and 
stepping probabilities generate statistical weights for chain conformations 
in a self-consistent field, and determine local volume fractions of connected 
segments. Flux balance across a unit lattice cell yields mean-field transport 
equations for free segment probabilities and momentum densities. Diffusive 
and viscous contributions to the fluxes arise from segmental hops modeled as 
a Markov process. Hopping transition rates depend on the changes in the free 
energy, reflecting segmental interactions, kinetic energy, and entropic 
contributions accounting for chain deformation under flow. It is the first 
dynamic mean-field theory for inhomogeneous polymer fluids that couples 
self-consistently the effects of flow, segmental and wall interactions, and 
compressibility on chain conformations, composition, transport and rheology 
in interfacial regions at the Kuhn length scale, which is necessary to 
resolve interfacial properties of typical polymer interfaces. This makes it 
different from other recently proposed approaches \cite{Maurits}. In this 
Letter, we focus on the interfacial dynamics of a symmetric polymer blend in 
the unentangled regime. We use the DSCF model to predict the velocity slip, 
chain conformations, and the reduction of shear viscosity across the 
interface between the coexisting polymer phases, and compare our results with 
molecular dynamics (MD) simulations.

We consider a blend of two immiscible homopolymer species A and B consisting 
of linear chains of $N^A$ and $N^B$ segments, respectively. The fluid is 
confined between two parallel walls normal to the $z$-axis in a Cartesian 
coordinate system at temperature $T$. The interface between the two polymers 
is parallel to, and is centered midway between the two walls. The walls move 
at equal and opposite velocities along the $x$-axis. We discretize the space 
between the two walls into triangular lattice layers parallel to the walls, 
stacked in a FCC lattice arrangement. The fluid system considered here is 
invariant under lattice translations in the $xy$-plane, and we simplify the 
DSCF equations utilizing this symmetry (see \cite{DSCF1} for details).

Let $P_i^\alpha$, $\phi_i^\alpha$ and $\mathbf{g}_i^\alpha$ be the free 
segment probability, segmental volume fraction and segmental momentum 
density of species $\alpha$ ($\alpha$=A or B) in each layer $i$ ($i$=$1, 2, 
\cdots, L$), respectively. The free segment probabilities and the segmental 
momentum density $\mathbf{g}_i$=$\mathbf{g}_i^A+\mathbf{g}_i^B$ evolve 
according to
\vspace{-0.1cm}
\begin{equation}
\frac{dP_i^\alpha}{dt}=-\grad\!\cdot\!(P_i^\alpha\mathbf{u}_i)
 -\left(\frac{\mathbf{j}_i^\alpha-\mathbf{j}_{i-1}^\alpha}{\sqrt{2/3}a}\right)
  \!\cdot \hat{\mathbf{z}}                                     \label{PiEqt}
\end{equation}
and
\begin{eqnarray}
\frac{d\mathbf{g}_i}{dt}
&=& \sum_{\alpha={\rm A},{\rm B}}\bigl[
       -\grad\!\cdot\!(\mathbf{g}_i^\alpha\mathbf{u}_i+\varepsilon^\alpha_i)
 -\left(\frac{\pi_i^\alpha-\pi_{i-1}^\alpha}{\sqrt{2/3}a}\right)
  \!\cdot \hat{\mathbf{z}}                             \nonumber \\
& &
       -\frac{\phi_i^\alpha}{w P_i^\alpha}
      (\zeta_{i,i+1}^\alpha\mathbf{j}_{i}^\alpha
         +\zeta_{i,i-1}^\alpha\mathbf{j}_{i-1}^\alpha)\bigr]~, 
                                                             \label{giEqt} 
\end{eqnarray}
which are obtained by applying mass and momentum conservation laws to a 
rectangular control volume of $w$=$a^3/\sqrt{2}$ centered at a site in layer 
$i$. Here $\zeta_{i,i\pm 1}^\alpha$ is the locally averaged segmental 
friction coefficient of species $\alpha$ (see \cite{DSCF1} for details) and 
$\mathbf{u}_i$=$\mathbf{g}_iw/(m^A\phi_i^A+m^B\phi_i^B)$, where $m^\alpha$ 
is the segmental mass of species $\alpha$, is the mass averaged velocity at 
a site in layer $i$. The tensor $-\varepsilon^\alpha_i$ in Eq.~(\ref{giEqt}) 
represents the elastic contribution to the stresss and is defined in 
\cite{DSCF1}. Let $D_{i,i+1}^\alpha$=$k_B T/N^\alpha\zeta_{i,i+1}^\alpha$ 
and $\nu_{i,i+1}^\alpha$=$\zeta_{i,i+1}^\alpha a^2/24 m^\alpha$ be the 
locally averaged self-diffusion coefficient and kinematic viscosity of 
species $\alpha$, and ${\bar \phi}$ be the overall average polymer volume 
fraction, the diffusive free segment probability current and the viscous 
stress tensor at the mid-plane between layers $i$ and $i+1$ in 
Eqs.~(\ref{PiEqt}) and (\ref{giEqt}) are given by
\begin{widetext}
\vspace{-0.67cm}
\begin{equation}
    \mathbf{j}_i^\alpha
= 3\sqrt{\frac{2}{3}}
   \frac{(1-\delta_{i+1,1})(1-\delta_{i,L})}{(1-{\bar \phi})}
   \frac{D_{i,i+1}^\alpha}{a}
  \left[P_i^\alpha(1-\phi_{i+1})
     \,\varphi\!\left(\frac{\Delta\! \left<H_i^\alpha\right>}{k_BT}\right)
        -P_{i+1}^\alpha(1-\phi_i)
   \,\varphi\!\left(\frac{-\Delta\! \left<H_i^\alpha\right>}{k_BT}\right)
  \right] \hat{\mathbf{z}}
                                            \label{DFlux}
\end{equation}
and
\begin{equation}
    \mathbf{\pi}_i^\alpha
= 3\sqrt{\frac{2}{3}}
   \frac{(1-\delta_{i+1,1})(1-\delta_{i,L})}{(1-{\bar \phi})}
   \frac{m^\alpha}{w} \frac{\nu_{i,i+1}^\alpha}{a}
  \left[\phi_i^\alpha \mathbf{u}_i (1-\phi_{i+1})
    \,\varphi\!\left(\frac{\Delta\! \left<H_i^\alpha\right>}{k_BT}\right)
        -\phi_{i+1}^\alpha \mathbf{u}_{i+1} (1-\phi_i)
   \,\varphi\!\left(\frac{-\Delta\! \left<H_i^\alpha\right>}{k_BT}\right)
  \right] \hat{\mathbf{z}}~.
                                            \label{VStress}
\end{equation}
\vspace{-0.2cm}
\end{widetext}
\vspace{-1cm}
In Eqs.~(\ref{DFlux}) and (\ref{VStress}), $\varphi$ is the Kawasaki 
transition rate function $\varphi(x)$=$2/(1+e^x)$ \cite{Kawasaki} which 
satisfies local detailed balance, and
$\Delta\!\left<H_i^\alpha\right>$=$\left<H_{i+1}^\alpha\right>-\left<H_i^\alpha\right>$ 
is the free energy change due to a hop of a segment of species $\alpha$ from 
layer $i$ to $i+1$. The local contribution to the free energy of a segment of 
species $\alpha$ in layer $i$ is
\begin{eqnarray}
    \left<H_i^\alpha\right>
&\!\!=\!\!& \sum_{\beta={\rm A},{\rm B}}\frac{(1-\delta_{\alpha\beta})}{2}
       \chi_{AB}\bigl<\!\bigl<\phi_i^\beta\bigr>\!\bigr>
      -\chi_s^\alpha(\delta_{i1}+\delta_{iL})             \nonumber  \\
& &  +\frac{m^\alpha}{2}\phi_i^\alpha\mathbf{u}_i^2
    -\frac{\phi_i^\alpha k_B T}{2N^\alpha}\left[
   {\rm Tr}\left(\mathbf{I}-\frac{3}{N^\alpha a^2}\mathbf{S}_i^\alpha
           \right) \right.                                \nonumber  \\
& & \left. +\ln\,\det\left(\frac{3}{N^\alpha a^2}
        \mathbf{S}_i^\alpha\right) \right]   \label{FreeEn}
\end{eqnarray}
where $\chi_{AB}$ is the segment-segment interaction parameter for species 
$A$ and $B$, and $\chi_s^\alpha$ is the segment-wall interaction parameter 
of a segment of species $\alpha$. In Eq.~(\ref{FreeEn}), the double angular 
brackets represent summation over all nearest neighbors of a site in layer 
$i$ and $\mathbf{S}_i^\alpha$ is the second moment of the end-to-end vector 
of an ideal (non-interacting) $\alpha$-type chain under flow, with its 
center-of-mass being in the $i$th layer. The last two terms in 
Eq.~(\ref{FreeEn}) are contributions from the kinetic energy and the free 
energy change due to stretching of the chains.

The conformations of interacting polymer chains are generated by lattice 
random walks in a self-consistent field modeling these interactions. 
According to \cite{SF}, the segmental volume fraction are obtained from the 
free segment probability by
\begin{equation}
  \phi_i^\alpha
= \frac{{\bar\phi}^\alpha L}{N^\alpha \sum_{j=1}^L P_j^\alpha(N^\alpha)}
     \sum_{s=1}^{N^\alpha} \frac{P_i^\alpha(s) P_i^\alpha(N^\alpha-s+1)}
                    {P_i^\alpha/(1-P_i^A-P_i^B)}  \label{phiI}
\end{equation}
where
\begin{eqnarray}
  P_i^\alpha(s)
&\!=\!& P_i^\alpha [\lambda_{+,i-1}^\alpha P_{i-1}^\alpha(s-1)
        +\lambda_{0,i}^\alpha P_i^\alpha(s-1)    \nonumber \\
&     & +\lambda_{-,i+1}^\alpha P_{i+1}^\alpha(s-1)]/(1-P_i^A-P_i^B)
                                             \label{recursive}
\end{eqnarray}
for $s=2, 3, \cdots , N^\alpha$ and 
$P_i^\alpha(1)=P_i^\alpha/(1-P_i^A-P_i^B)$. In Eq.~(\ref{phiI}), 
${\bar\phi}^\alpha$ is the average volume fraction of species $\alpha$ in 
the system and in Eq.~(\ref{recursive}), $\lambda_{\pm,i}^\alpha$ is the 
stepping probability of a $\alpha$-type segments in an ideal chain from 
layer $i$ to layer $i\pm 1$, whereas $\lambda_{0,i}^\alpha$ is the stepping 
probability from a site in layer $i$ to a nearest neighbor in the same 
layer. In contrast to the static SCF theory, these probabilities are now 
time-dependent and anisotropic, and are related to the components of 
$\mathbf{S}_i^\alpha $ through the random walk picture (see \cite{DSCF1}).

Finally, $\mathbf{S}_i^\alpha $ evolves according to
\begin{eqnarray}
& &  \frac{D\mathbf{S}_i^{\alpha}}{Dt}
         -(\grad \mathbf{u}_i)^{\rm T}\!\cdot\mathbf{S}_i^\alpha
         -\mathbf{S}_i^\alpha\!\cdot(\grad \mathbf{u}_i) \nonumber  \\
&= &   -\frac{1}{\tau_{db,i}^\alpha}
    \left[
       \frac{\mathbf{S}_i^\alpha}{1-\frac{1}
              {N^\alpha(N^\alpha-1) a^2}{\rm Tr}\,\mathbf{S}_i^\alpha}
      -\frac{N^\alpha a^2}{3}\mathbf{I}
    \right]~,                  \label{stressEqt}
\end{eqnarray}
which is the constitutive equation of the FENE-P dumbbell model 
\cite{FENEP}, but with the value of the spring constant chosen in such a way 
that $\mathbf{S}_i^\alpha$ reverts to the Hookean dumbbell value in 
equilibrium. In Eq.~(\ref{stressEqt}), 
$\tau_{db,i}^\alpha$=${N^\alpha}^2\zeta_i^\alpha a^2/24 k_B T$ is the local 
dumbbell relaxation time where $\zeta_i^\alpha$ is the local segmental 
friction coefficient of species $\alpha$, which is related to the one in the 
bulk through the Doolittles' law \cite{Doolittle} (see \cite{DSCF1} for 
details).

In the actual computation, all spatial derivatives are approximated by using 
finite difference formulas. Under suitable boundary conditions, 
Eqs.~(\ref{PiEqt}), (\ref{giEqt}) and (\ref{stressEqt}) form a closed system 
of ordinary differential equations, which is solved numerically.

Henceforth we focus on symmetric blends, i.e. $N^A\!=\!N^B\!=\!N$, 
$m^A\!=\!m^B\!=\!m$, and the two polymers A and B have the same density and 
segmental friction coefficient. We also set $\chi_s^A\!=\!\chi_s^B\!=\!0$. 
We shall compare the numerical solutions of the DSCF equations with 
Molecular Dynamics (MD) simulations of a similar system. The details of the 
model and the method of the MD simulation that we used can be found in 
\cite{MDRef}. In MD, the characteristic energy ($\epsilon$), time ($\tau$) 
and length ($\sigma$), and the segmental mass ($m$) are related by 
$\tau=\sigma\sqrt{m/\epsilon}$. As in \cite{MDRef}, we set the temperature 
in our MD simulations by $\epsilon=k_BT/1.1$.

In order to compare our model to MD simulations, we have to establish the 
relationships between the model parameters in the DSCF theory and those in 
MD. The Kuhn length in the MD simulations is estimated to be $a\!\approx\! 
1.3\sigma$ \cite{MDRef}, hence, one Kuhn segment is equivalent to 1.3 beads 
in MD. The interaction parameter $\chi_{AB}$ is obtained by fitting the 
equilibrium density profiles calculated from DSCF theory to the ones 
obtained from MD. After accounting for capillary wave broadening of the 
intrinsic profiles in MD simulations \cite{MDprofile}, this gives 
$12\chi_{AB}/k_B T \!\approx\! 1.1$. The segmental friction coefficient in 
the bulk is obtained by comparing the self-diffusion constants calculated 
from MD to the one in our model. The segmental mass in our model 
is identical to the one used in MD. In the simulations, we set 
${\bar\phi}^A\!=\!{\bar\phi}^B$ and the overall polymer volume fraction 
averaged over the simulation box to 
${\bar\phi}={\bar\phi}^A\!+{\bar\phi}^B\!=0.85$. This average volume 
fraction is used both in the DSCF model and MD simulations.

As the first test of the DSCF model, we calculated the steady state velocity 
profiles and the local viscosities between the walls under shear for 
different chain lengths, and compared them with the MD results. The steady 
state is obtained by integrating the DSCF equations of motion in time under a 
constant rate for shearing the two walls (nominal shear rate), starting from 
equilibrium.

\begin{figure}
\begin{center}
  \includegraphics[angle=-90,width=0.48\textwidth]{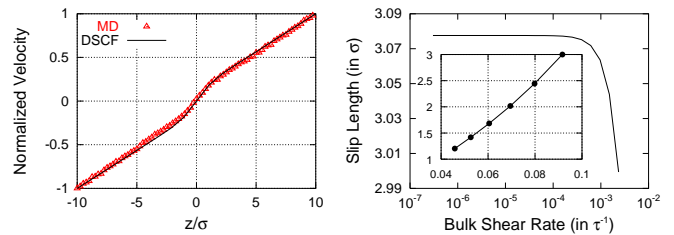}
\end{center}
\vspace{-0.6cm}
\caption{\label{VelProfile} Left: Steady state shear velocity profile for 
$N$=12 obtained from DSCF at $\dot\gamma$=2.36$\times$$10^{-3}\tau^{-1}$ and 
MD at $\dot\gamma$=4.04$\times$$10^{-3}\tau^{-1}$. (The velocity is 
normalized by the velocity at $z=10\sigma$.) Right: Variation of slip length 
with shear rate and with $\chi_{AB}$ for $N$=12 obtained from DSCF. (Inset 
shows the slip length in $\sigma$ Vs $\chi_{AB}/k_B T$ at nominal shear rate 
2.75$\times$$10^{-3}\tau^{-1}$.)}
\end{figure}

The shear velocity profile (defined as the local mass-averaged velocity) 
across the polymer-polymer interface in the middle of the channel from our 
model for $N$=12 is shown in the left panel of Fig.~\ref{VelProfile}. The 
result from MD simulations using chains of 16 beads is also shown. The shear 
rates $\dot\gamma$ in the bulk are indicated in the caption \cite{MD-Note}.  
This bulk shear rate is obtained by fitting the linear part of the velocity 
profile to a straight line and is different from the nominal shear rate 
because of the velocity slip at the walls and the interface. Note that a kink 
representing a velocity slip at the interface is observed here. This 
interfacial slip phenomenon was first predicted by de Gennes \cite{deGennes} 
based on a scaling argument, and later modeled with an approximate 
constitutive equation for incompressible polymer blends by Goveas and 
Fredrickson \cite{Fredrickson}, and was subsequently confirmed by Barsky and 
Robbins \cite{BarskyRobbins1} using MD. Experimental studies have also 
indicated the presence of a velocity slip at the interface in polymer blends 
\cite{Macosko}. Here the same velocity slip is produced in our DSCF model. 
The bulk shear rates for the DSCF and MD results shown in 
Fig.~\ref{VelProfile} are not identical, but the closest that we can attain. 
This is because simulating systems with low shear rates in MD is not 
computationally feasible. On the other hand, the DSCF simulations currently 
become unstable at higher shear rates, though they cover most experimentally 
accessible shear rates.

To quantify the interfacial slip, we calculated the slip length, which is 
defined as twice the magnitude of the $x$-intercept of the straight line 
obtained from a linear fit to the linear part of the steady state velocity 
profile. The results are shown in the right panel of Fig.~\ref{VelProfile}.  
Our model shows that the slip length decreases as the shear rate increases. 
We verified this result by performing MD simulations of the same system at 
two different bulk shear rates: one at 4.04$\times$$10^{-3}\tau^{-1}$ and 
the other one at 8.27$\times$$10^{-3}\tau^{-1}$. The slip lengths were found 
to be $1.95\sigma$ and $1.22\sigma$, respectively. The trend is the same as 
the one predicted by the DSCF model. We also calculated the slip length as a 
function of $\chi_{AB}$ at a fixed nominal shear rate and the result is 
shown in the inset of the same figure. It shows that the slip length 
increases as the strength of the segment-segment interaction increases. A 
similar qualitative trend was also observed by using MD in earlier studies 
\cite{BarskyRobbins1}.

\begin{figure}
\begin{center}
  \includegraphics[angle=-90,width=0.48\textwidth]{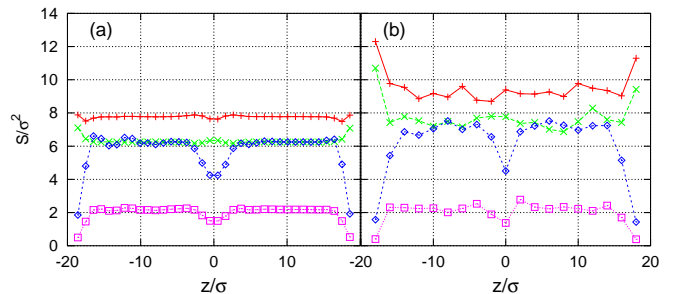}
\end{center}
\vspace{-0.6cm}
\caption{\label{QQ} Second moment of the end-to-end distance for $N$=12 
obtained from (a) DSCF at $\dot\gamma$=2.36$\times$$10^{-3}\tau^{-1}$, and 
(b) MD at $\dot\gamma$=4.04$\times$$10^{-3}\tau^{-1}$. (+: $S_{xx}$, 
$\times$: $S_{yy}$, {\Large $\diamond$}: $S_{zz}$, $\Box$: $S_{xz}$.)}
\end{figure}

\begin{figure}
\begin{center}
  \includegraphics[angle=-90,width=0.48\textwidth]{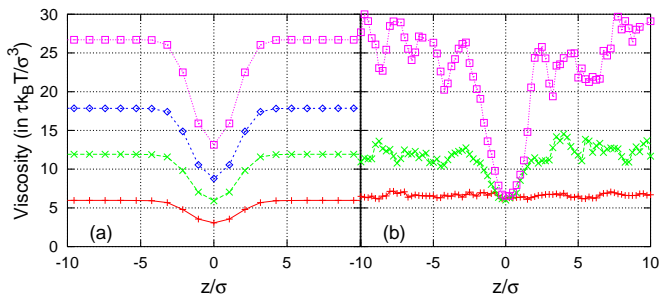}
\end{center}
\vspace{-0.6cm}
\caption{\label{VisProfile} Steady state viscosity profiles across the 
polymer-polymer interface for different chain lengths obtained from (a) DSCF 
at nominal shear rate 6.24$\times$$10^{-4}\tau^{-1}$, and (b) MD at various 
nominal shear rates. (+: $N$=6, $\times$: $N$=12, {\Large $\diamond$}: 
$N$=18, $\Box$: $N$=24.)}
\end{figure}

We also calculated the second moment of the end-to-end vector at steady 
state by performing a Monte-Carlo simulation of lattice random walks in a 
self-consistent field representing conformations of interacting chains under 
shear. The transition rates used in the Monte-Carlo simulation were 
constructed from the stepping and free segment probabilities from the DSCF 
theory. The result is shown in Fig.~\ref{QQ} and compared to the one 
obtained from MD. In the bulk, we have good agreement between the two 
simulations. The slight difference can be accounted for by the difference in 
the shear rates. In the interfacial region, our model predicts the same 
trend as in MD. Similar results were also obtained by Barsky and Robbins in 
\cite{BarskyRobbins2} using MD.

The steady state viscosity profiles in the interfacial region for various 
chain lengths are calculated by dividing the viscous shear stress by the 
local shear rate, and are shown in Fig.~\ref{VisProfile}(a). Note that a 
drop in the viscosity occurs at the polymer-polymer interface and it is more 
pronounced for longer chains. Corresponding results from MD are shown in 
Fig.~\ref{VisProfile}(b). Here one can see that the trend predicted by the 
DSCF model is in general agreement with the MD results. The shear 
viscosities obtained from both methods in the bulk regions are very close. 
In the interfacial region, the drop is underestimated for long chains and 
slightly overestimated for short chains in the DSCF model, but the general 
agreement is very good given all the approximations of the model. A similar 
drop in the viscosity was also observed by Barsky and Robbins using MD 
\cite{BarskyRobbins1}, but with different parameters.

We have studied the interfacial slip phenomenon in phase separated sheared 
polymer blends using the novel DSCF theory and compared our results with MD 
simulations. Good agreement between the two approaches is found. For 
unentangled polymer fluids studied here, the DSCF simulation time is only a 
small fraction of that in MD. We expect that these computational efficiences 
will be more manifest in the case of entangled polymer fluids, for which MD 
simulations are currently impractical. The current DSCF theory is based on 
conservation laws for species occupancies and momentum that are coupled to 
models of polymer structure and conformation. It can be generalized either 
by changing the constitutive stress equation (accounting for the reptation 
regime), the polymer model (e.g., accounting for block copolymers or 
branching) or by incorporating additional conservation laws (e.g., for 
energy and/or charge). Work along these lines will be reported in future 
publications.

This work was supported by the Mitsubishi Chemical Corporation of Japan. M. 
Mihajlovic and Y. Shnidman would also like to acknowledge support by a grant 
from the National Science Foundation (DMR-0080604). We would like to 
acknowledge useful discussions with Drs. G. Fredrickson and D. Wu. We thank 
Dr. T. Kawakatsu for sending his preprint prior to its publication. The work 
of the first three authors was performed at Polytechnic University in 
Brooklyn.


\begin{thebibliography}{99}
\bibitem{SF} J. Scheutjens and G.J. Fleer, J. Phys. Chem. {\bf 83}, 1619
(1979).

\bibitem{DSCF1} M. Mihajlovic, T.S. Lo and Y. Shnidman. (submitted to 
Macromolecules); M. Mihajlovic, Ph.D. thesis, Polytechnic University (2004).

\bibitem{CDLG} A.A. Khan and Y. Shnidman, Progr. Colloid. Polym. Sci. {\bf
103}, 251 (1997).

\bibitem{Maurits} N.M. Maurits, A.V. Zvelindovsky and J. Fraaije, J. Chem. 
Phys. {\bf 109}, 11032 (1998); G.H. Fredrickson, J. Chem. Phys. {\bf 117}, 
6810 (2002); T. Shima, H. Kuni, Y. Okabe, M. Doi, H.F. Yuan and T.
Kawakatsu, Macromolecules {\bf 36}, 9199 (2003).

\bibitem{Kawasaki}B. Schmittmann and R.K.P. Zia, {\em Statistical Mechanics 
of Driven Diffusive Systems}. (Academic Press, London, 1995).

\bibitem{FENEP} R. Bird, C. Curtiss, R. Armstrong and O. Hassager, {\em 
Dynamics of Polymeric Liquids, 2nd Ed, Vol. 2}. (John Wiley \& Sons, 1987); 
M. Herrchan and H.C. Ottinger, J. Non-Newtonian Fluid Mech. {\bf 68}, 17 
(1997).

\bibitem{Doolittle} A.K. Doolittle and D.B. Doolittle, J. Appl. Phys. {\bf
28}, 901 (1957).

\bibitem{MDRef} K. Kremer and G. Grest, J. Chem. Phys. {\bf 92}, 5057
(1990).

\bibitem{MDprofile} A.N. Semenov, Macromolecules {\bf 27}, 2732 (1994); M.D. 
Lacasse, G.S. Grest and A.J. Levine, Phys. Rev. Lett. {\bf 80}, 309 (1998); 
K. Binder, M. Muller, F. Schmid and A. Werner, Adv. Coll. Interf. Sci. {\bf
94}, 237 (2001).

\bibitem{MD-Note} In comparing DSCF results with the MD simulations, we made 
sure that the MD simulations were operating in a range in which the shear 
stress is independent of shear rate. This allows us to compare the two
methods, even though the shear rates are not identical.

\bibitem{deGennes} P.G. de Gennes, C.R. Acad. Sci. {\bf 308 II}, 1401 
(1989); F. Brochard, P.G. de Gennes and S. Troian, C.R. Acad. Sci. Paris 
{\bf 310 III}, 1169 (1990).

\bibitem{Fredrickson} J.L. Goveas and G.H. Fredrickson, Eur. Phys. J. B {\bf 
2}, 79 (1998).

\bibitem{BarskyRobbins1} S. Barsky and M.O. Robbins, Phys. Rev. E, {\bf 63}, 
021801 (2001).

\bibitem{Macosko} R. Zhao and C.W. Macosko, J. Rheol., {\bf 46}(1), 145 
(2002).

\bibitem{BarskyRobbins2} S. Barsky and M.O. Robbins, Phys. Rev. E, {\bf 65}, 
021808 (2002).
\end{thebibliography}
\end{document}